\documentclass[nofootinbib,twocolumn,superscriptaddress]{revtex4}
\usepackage{amsmath}
\usepackage{amsthm}
\usepackage{mathtools}
\usepackage{amssymb}
\usepackage{dsfont}
\usepackage{subfigure}
\usepackage{upgreek}
\usepackage[english]{babel}
\usepackage{siunitx}
\begin{document}

\title{Test of macroscopic realism with coherent light}

\author{Hui Wang}
\affiliation{School of Physical Sciences \& Key Laboratory of Vacuum Physics, \\ University of Chinese Academy of Sciences, YuQuan Road 19A, Beijing 100049, China}
\affiliation{International Centre for Theoretical Physics Asia-Pacific, University of Chinese Academy of Sciences, Beijing 100190, China}

\author{Shuang Wang}
\affiliation{School of Physical Sciences \& Key Laboratory of Vacuum Physics, \\ University of Chinese Academy of Sciences, YuQuan Road 19A, Beijing 100049, China}

\author{Cong-Feng Qiao}
\email{qiaocf@ucas.ac.cn}
\affiliation{School of Physical Sciences \& Key Laboratory of Vacuum Physics, \\ University of Chinese Academy of Sciences, YuQuan Road 19A, Beijing 100049, China}

\begin{abstract}

    Macro-realism is a fundamental feature of  classical world that contradicts with the quantum theory. An elegant method  of testing macrorealism is to apply the Leggett-Garg inequality (LGI), but the non-invasivity of measurement is challenging in practice. In this work, we report the LGI violation of  path observable in a composite interference experiment with coherent light. Experiment results confirm the occurrence of destructive interference providing per se as evidence of macro-realism violation. And by using an exact weak measurement model in the present experiment,  the advantage that the violation of realism is independent of the invasive strength allows the realization of direct measurement and strengthens the persuasion of verification at the macroscopic scale.

\end{abstract}

\maketitle

\section{introduction}

In 1985, Leggett and Garg  provided an elegant method to experimentally test macrorealism with  a class of inequalities \cite{Lg1985}. The Leggett-Garg inequalities
(LGIs) are based on the following two assumptions.
(1) Macrorealism per se (MRps): A macroscopic system always remains in one of its available two or more  distinct ontic states.
(2)
Noninvasive measurability (NIM): In principle, it is possible to determine the state of  macrosystem  without disturbing it.
The classical world  we daily  experience   intuitively obeys these assumptions, while quantum mechanics is not satisfied because of the superposition principle and the collapse of  state. Hence the LGIs act as an indicator of  the existence of coherence, and test the applicability of quantum mechanics when scaling from microscopic to macroscopic systems \cite{Leggett2002}.

Considering  dichotomic observable $\hat {M}=\pm1$ in  macrorealistic system, the  standard LGI for the $n$-time measurement  is given by \cite{ntime}
\begin{equation}\label{ntim}
 K_{n}=\left\langle\hat{M}_{1} \hat{M}_{2}\right\rangle+\left\langle\hat{M}_{2} \hat{M}_{3}\right\rangle + \cdots +\left\langle\hat{M}_{n-1} \hat{M}_{n}\right\rangle-\left\langle\hat{M}_{1} \hat{M}_{n}\right\rangle .
\end{equation}
Here, $ \langle \hat{M}_{i} \hat{M}_{j}\rangle=\sum_{m_{i}, m_{j}} m_{i} m_{j} P\left(M_{i}^{m_{i}}, M_{j}^{m_{j}}\right)
 $ is the correlation function, where $P\left(M_{i}^{m_{i}}, M_{j}^{m_{j}}\right)$  is the joint probability of  obtaining
 measurement  $m_{i}$ and $m_{j}$  at sequential time $t_{i}$ and $t_{j}$.
 The bound of  inequality (\ref{ntim})  is formulated as  follows \cite{LGbound}. If $n$ is odd, $-n \leqslant K_n  \leqslant n-2$, and if $n$ is even, $-(n-2) \leqslant K_n \leqslant n-$ 2. For $n=3$,  the typical
 LGI is written as
 \begin{equation}\label{o}
    K_{3}=\left\langle\hat{M}_{1} \hat{M}_{2}\right\rangle + \left\langle\hat{M}_{2} \hat{M}_{3}\right\rangle-\left\langle\hat{M}_{1} \hat{M}_{3}\right\rangle ,
\end{equation}
and  satisfies $-3 \leqslant K_3  \leqslant 1$. However, the LGI will go beyond
 bound if the system is quantum. For a two-level system,
  the maximum quantum value of $n$-time case   is $ (K_{n})_{Q} 
  = n \cos \frac{\pi}{n}$,  which is
  referred to as the L\"uders  bound \cite{QMbound}.  For $ K_3$ and $K_4$  the L\"uders bound are  $1.5$ and $2\sqrt{2}$, respectively.    Recently, it has been predicted that the quantum violation value can exceed  L\"uders  bound. One is to relax the measurement assumption from the L\"uders state-update rule to a general projective measurement for multilevel system, e.g.,  $K_{3}^{\text{max}}=1.7565$ in three-level system \cite{Budroni2014}. The other is the new variant of LGI with three-time, two-time, and  one-time correlation function, where $K_{3}^{\text{max}}=1.93$ \cite{Akpan}.

  In terms of experiment, the primary and decisive condition for verifying macroscopic realism is the implementation of  NIM  assumption, which is also the  main challenge in practice. Specific methods include  ideal negative
result measurement (INRM) \cite{INRM} and
weak measurement \cite{WWA,SemiWWA}. The former  confirm that there is no interaction between the measured system and the auxiliary  instrument by keeping  only the negative results. The latter, in principle, is capable of  imposing an
extrem-weak disturbance on system state.
Up to now, LGI has been experimentally tested in a variety of microsystems, i.e. nuclear spins \cite{NMR2011,NMR2013,NMR2017}, nitrogen-vacancy center \cite{Du2022}, single photons \cite{xue2018,xue2020,PRXQuantum2022}, superconducting flux qubit \cite{supd},  and light-matter interaction \cite{ggc2015}.
The  results indicate the  violations of LGI  for each individual quantum measurement, but it does not sufficiently infer that it also holds in macroscopic system, e.g., an ensemble of photons.

With the exception of quantum system, the LGI gains new focus  on classical optical system \cite{classical2019,classical2021}. Experiments  investigated the violation of LGI by using the polarization  freedom degree of the laser beam \cite{c2021}. However, the lack of quantification of  invasiveness introduced by the  measurement instrument is an obvious  loophole and  fails to satisfy the assumption of NIM. In our work, this loophole is closed by exploiting weak measurement technique.

Recently, A. K. Pan. proposed a new correlation  between interference experiment and the  violation of LGI with the assistance of  anomalous weak value, which suggested the avoidance of the NIM assumption \cite{Pan2020}. However, achieving it in experiment is hardly available. Here we tested this LGI variant in  the path interference experiment of classical light. Results show   the existence of anomalous weak value is alway accompanied by destructive interference, and consequently, LGI is identified to be violated in this range.
Differ from the previous ones, violation of macro-realism is inferred straight from the destructive interference per se. Therefore,  we implement a
direct test  for the macroscopic realism by using precise weak measurement.

\section{theory and methodology}

In this section, we prior to review the derivation in Ref. \cite{Pan2020}.
The LGI for the three-time measurement scenario
can also be expressed as
\begin{equation}\label{orgin}
    \begin{aligned}
    K_{3}= & 1-m_{1}m_{2}\left\langle\hat{M}_{1} \hat{M}_{2}\right\rangle-m_{2}m_{3}\left\langle\hat{M}_{2} \hat{M}_{3}\right\rangle  \\ & +m_{1}m_{3}\left\langle\hat{M}_{1} \hat{M}_{3}\right\rangle \geqslant 0 ,
\end{aligned}
\end{equation}
where $m_{i}m_{j}=\pm 1 (i,j=1,2,3) $. Specially,  $m_{1}=m_{2}=m_{3}= 1$, one simply recovers inequality (\ref{o}).

The measurement  $\hat{M}_{1}$ at $t_{1}$ directly
satisfy $\hat{M}_{1}\left|+m_{1}\right\rangle=\left|+m_{1}\right\rangle$ and  $m_{1}=+1$ when the initial system  is prepared  at particular    state $\left|\psi_{i}\right\rangle=\left|+m_{1}\right\rangle$.
Hence for different values of
 $m_{2}=\pm 1$ and $m_{3}=\pm 1$, Eq. (\ref{orgin}) evolved into four two-time LGIs
 \begin{equation}\label{orgin1}
\begin{aligned}
K_{31} &=1-\left\langle\hat{M}_{2}\right\rangle-\left\langle\hat{M}_{2} \hat{M}_{3}\right\rangle+\left\langle\hat{M}_{3}\right\rangle \geqslant 0, \\
K_{32} &=1+\left\langle\hat{M}_{2}\right\rangle+\left\langle\hat{M}_{2} \hat{M}_{3}\right\rangle+\left\langle\hat{M}_{3}\right\rangle \geqslant 0, \\
 K_{32}&=1-\left\langle\hat{M}_{2}\right\rangle+\left\langle\hat{M}_{2} \hat{M}_{3}\right\rangle-\left\langle\hat{M}_{3}\right\rangle \geqslant 0, \\
K_{34} &=1+\left\langle\hat{M}_{2}\right\rangle-\left\langle\hat{M}_{2} \hat{M}_{3}\right\rangle-\left\langle\hat{M}_{3}\right\rangle \geqslant 0 .
\end{aligned}
\end{equation}
In fact, there is only one  output per measurement, so  inequalities (\ref{orgin1})  are impossible to hold at the same time.  Mathematically articulated as
$\hat{M}_{3}=2\left|+m_{3}\right\rangle\left\langle+m_{3}|-\mathbb{I}=\mathbb{I}-2|-m_{3}\right\rangle\left\langle-m_{3}\right|$.
So,  the projective measurement  of $\hat{M}_{3}$ at time $t_{3}$ is performed by $\left|\psi_{f}\right\rangle=\left|+m_{3}\right\rangle$
or $\left|-m_{3}\right\rangle$.   The process of setting a fixed option at initial and last time, in essence, provide  pre- and post-selective state for weak measurement. Under the constraint  of  a two-state vector,
the left  sides of above four LGIs  are  manipulate as
\begin{equation}\label{orgin2}
\begin{aligned}
&\left(K_{31}\right)_{Q}=2 p_{|+m_{3}\rangle}\left[1-\left\langle\hat{M}_{2}\right\rangle_{w}^{\left.\mid+m_{3}\right\rangle}\right], \\
&\left(K_{32}\right)_{Q}=2 p_{|+m_{3}\rangle}\left[1+\left\langle\hat{M}_{2}\right\rangle_{w}^{\left.\mid+m_{3}\right\rangle}\right], \\
&\left(K_{33}\right)_{Q}=2 p_{|-m_{3}\rangle}\left[1-\left\langle\hat{M}_{2}\right\rangle_{w}^{\left.\mid-m_{3}\right\rangle}\right], \\
&\left(K_{34}\right)_{Q}=2 p_{|-m_{3}\rangle}\left[1+\left\langle\hat{M}_{2}\right\rangle_{w}^{\left.\mid-m_{3}\right\rangle}\right],
\end{aligned}
\end{equation}
where
$\langle\hat{M}_{2}\rangle_{w}^{\left|\pm m_{3}\right\rangle}=\frac{\left\langle \pm m_{3}\left|\hat{M}_{2}\right| + m_{1}\right\rangle}{\left\langle \pm m_{3} \mid  + m_{1}\right\rangle}$ is  the weak value of $\hat{M}_{2} $  and
$p_{|\pm m_{3}\rangle}=|\left\langle \pm m_{3}   \mid  +m_{1}   \right\rangle|^2$  is  the postselection
probabilitiy.
The  formula (\ref{orgin2}) contains the association between LGI, weak value and postselection probability.
It can be inferred that the LGI will be violated that $(K_{3i})_{Q} < 0$ when the weak value  exceed  eigenvalue range  $ [-1, 1]$.

There is an emphasis   on path-only interference experiment in Ref. \cite{Pan2020}. That is, the NIM assumption is dispensable because no additional measurement apparatus is introduced.  In the Mach-Zehender (MZ) set up,
the  initial system  state is
$\left|\psi_{i}\right\rangle=\cos \theta \left|U\right\rangle + \sin \theta \left|D\right\rangle$, where  $\{\left|U\right\rangle, \left|D\right\rangle \}$ are orthogonal path states.
And the projected states are $\left|+ m_{3}\right\rangle=\left(\left|U \right\rangle+\left|D \right\rangle\right) / \sqrt{2} $, $\left|- m_{3}\right\rangle=\left(\left|U\right\rangle-\left|D\right\rangle\right) / \sqrt{2} $  with the probabilities of $p_{|+m_{3}\rangle}=(1+\sin 2\theta) /2$ and
$p_{|-m_{3}\rangle}=(1-\sin 2\theta ) /2$, respectively. 
Thus, for the dichotomic  observable $\hat M_{2}=|U\rangle \langle U|-|D\rangle \langle D|$,  a simple substitution  Eq. (\ref{orgin2}) turns to
\begin{equation}\label{x2}
\begin{aligned}
&\left(K_{31}\right)_{Q}=2\sin \theta(\cos \theta + \sin \theta) ,\\
&\left(K_{32}\right)_{Q}=2\cos \theta(\cos \theta + \sin \theta),\\
&\left(K_{33}\right)_{Q}=2\sin \theta( \sin \theta -\cos \theta ), \\
&\left(K_{34}\right)_{Q}=2\cos \theta(  \cos \theta -\sin \theta).
\end{aligned}
\end{equation}
Here, weak value
$\langle\hat{M}_{2}\rangle_{w}^{|+m_{3}\rangle}= (\cos \theta- \sin \theta )/ (\cos \theta + \sin \theta)$ and
$\langle\hat{M}_{2}\rangle_{w}^{|- m_{3}\rangle}=( \cos \theta + \sin \theta)/(\cos \theta- \sin \theta)$ are real.

The  path-only scheme is logical in theory, but for this procedure to
work, one has to consider the setting of  weak disturbance and  variable $\theta$.
In this context, we adopt the exact weak  measurement method  and   regulate parameter $\theta$ with the polarization degrees of freedom \cite{chenjs}. A particularly notable feature of this experimental design is that
 the weak value of Pauli type observable is independent of the coupling strength. It is further  deduced that the results are consistent whether the perturbation is present or absent.
This effect is similar to the case of path-only interference, i.e., avoiding the use of NIM.
 In other words, the test for macroscopic realism  can be done directly.

\begin{figure} [h]\centering
    \includegraphics[width=0.5\textwidth]{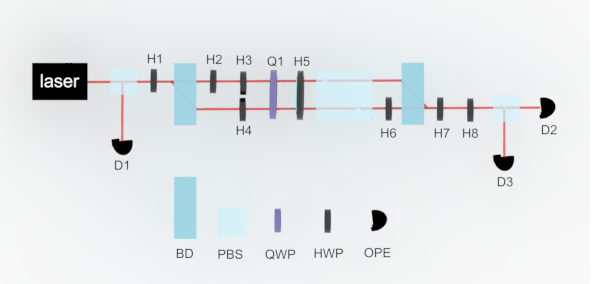}
    \caption{Experimental setup. The laser is prepared to the state $\left|\psi\right\rangle=\cos \theta \left|H\right\rangle + \sin \theta \left|V\right\rangle$ through a polarizing beam splitter (PBS) and half
    wave plate (H1). Then it enters an interference network consists of two polarizing beam displacers (BD), where polarization and path degrees of freedom act as pointer and measured system. Sets of H3 and H4 on the upper and lower path are used to realize the weak interaction $\hat{ \mathbf {U} }$. Projective
    measurement of pointer  is performed via Q1, H5, and  PBS. Finally, the multiple waveplates (H6-8), BD  and  PBS are applied to generate the post-selection state $\left|\psi_{f}\right\rangle=\left|\pm m_{3}\right\rangle$. }\label{f1}
\end{figure}

Following the above  motivation,  the experimental setup  is implemented by  a controlled interference  network as depicted in Fig. \ref{f1}. The optical interference   consists of two polarizing beam displacers (BDs)  and multiple waveplates.
BDs can  transmit vertical polarized light directly and  deflect horizontal light laterally by the displacement of \SI{4}{mm}.
The laser beam   passes through  polarizing beam splitter (PBS) and  followed  half wave plate (HWP, H1)  with the  setting angle $\theta /2$  in the transmission path, providing the  linearly polarized state $\left|\psi\right\rangle=\cos \theta \left|H\right\rangle + \sin \theta \left|V\right\rangle$. The reflection path of PBS serves as a reference and records the fluctuation of light intensity.
 The first BD splits the  state $\left|\psi\right\rangle$ into  two parallel spatial modes, upper mode and lower mode encoded as   $|U\rangle$ and  $|D\rangle$, respectively. Then  the polarization information $\{\left|H\right\rangle, \left|V\right\rangle \}$ transform into $\{\left|U\right\rangle, \left|D\right\rangle \}$ via the H2 rotated at $\pi/4$ in the upper path. The pre-selected state  $\left|\psi_{i}\right\rangle$ and pointer state $|H\rangle$ are prepared. So the  composite state combined with the pointer and measured system is clearly expressed as  $|H\rangle(\cos \theta \left|U\right\rangle + \sin \theta \left|D\right\rangle) $ at the initial time.
Subsequently, the    weak interaction $\hat{ \mathbf {U} }= e^{-i \gamma \hat{M}_{2} \otimes \hat{\sigma}_{y}}$ is realized by placing H3( $\gamma /2$) and H4( $-\gamma /2$)   on the upper and lower path, respectively.

Following the preparation  method of pre-selected state, the post-selection state is done  through  BD2, HWPs (H6-8)  and PBS, where  H6(H7)  and  H8 are rotated at $\pi/4$ and  $\pi/8$  radians, respectively. The detection ports D2 and D3 correspond to the  projection basis of  system
$\left|+ m_{3}\right\rangle$ and $\left|- m_{3}\right\rangle$.  As a result, the information of the measured system (weak value) is captured in the pointer state. More explicitly, the relation between  weak value  $\langle\hat{M}_{2}\rangle_{w}$ and  the expectation value of pointer  observable $\left\langle\hat{\sigma}_{i}\right\rangle_{p}$
 is exactly expressed as \cite{chenjs}
\begin{equation}\label{xx2}
    \begin{aligned}
    \left\langle\hat{\sigma}_{x}\right\rangle_{p}  =\frac{\sin 2 \gamma \operatorname{Re}\langle \hat M_{2}\rangle_{w} }{\cos ^{2}\gamma+\sin ^{2} \gamma |\langle\hat M_{2} \rangle_{w}|^{2}},
    \\
    \left\langle\hat{\sigma}_{y}\right\rangle_{p} =\frac{\sin (2 \gamma) \operatorname{Im}\langle \hat M_{2}\rangle_{w} }{\cos ^{2}\gamma+\sin ^{2} \gamma |\langle\hat M_{2} \rangle_{w}|^{2}}.
    \end{aligned}
    \end{equation}

 According to the formula (\ref{xx2}), the real and imaginary part of  weak value can be extracted from
\begin{equation}\label{s}
 \begin{aligned}
    \left\langle\hat{\sigma}_{x}\right\rangle_{p}=\frac{(I_{+}-I_{-})}{(I_{+}+I_{-}) }  \qquad
    \left\langle\hat{\sigma}_{y}\right\rangle_{p}=\frac{(I_{R}-I_{L})}{(I_{R}+I_{L})},
 \end{aligned}
\end{equation}
where the  intensity at the corresponding polarization is denoted by $I_{i}$, the  polarization
 $ |\pm\rangle = \frac{1}{\sqrt{2}}(|H\rangle \pm |V\rangle)$ and $ |L (R)\rangle = \frac{1}{\sqrt{2}}(|H\rangle \pm i|V\rangle).$
 So, the  quarter-wave plate  (Q1), H5, and  PBS are set to realize
 the projection measurement of  pointer  $\{|+\rangle,|-\rangle\}$ and $\{|R\rangle,|L\rangle\}$.
 Note that  the projection measurement was preferentially performed in the experiment setup, because it is   physically commutable with the post-selection of measured system. It does not affect the measurement result for which one  is performed in priority.

\section{results}

\begin{figure}[h] \centering
    \includegraphics[width=0.5\textwidth]{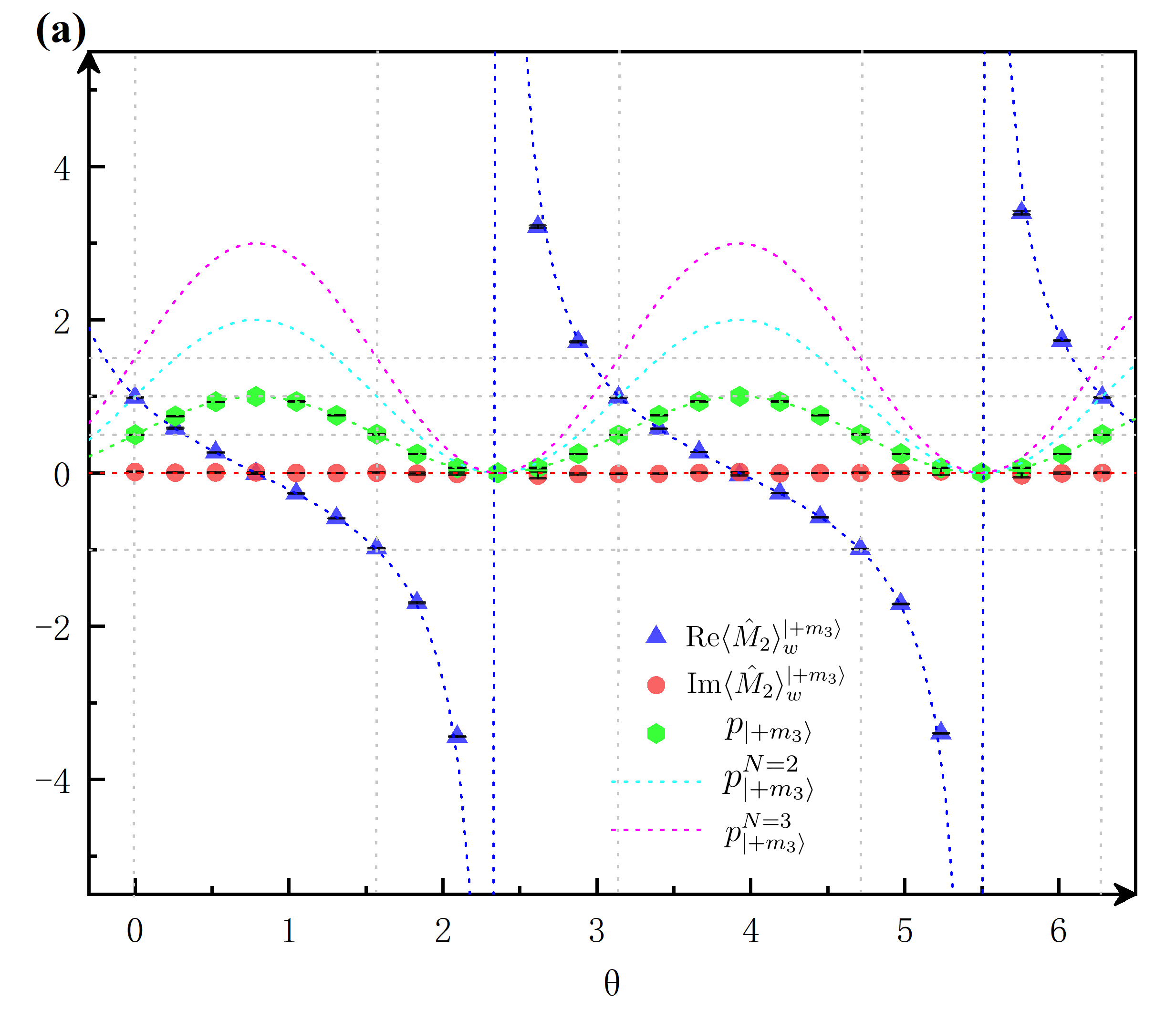}
    \includegraphics[width=0.5\textwidth]{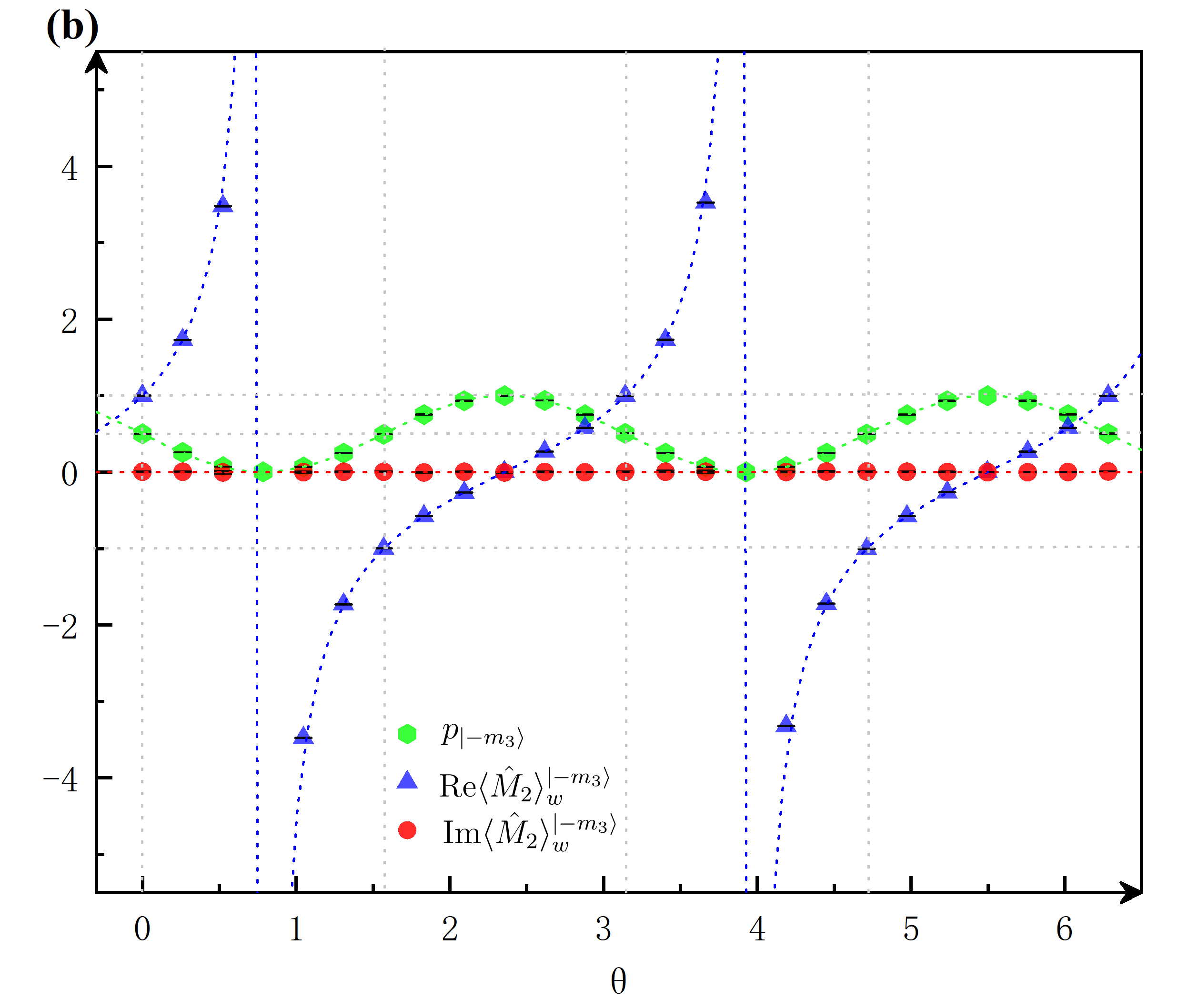}
    \caption{Experimental results of  the post-selection probability $p$ and weak value $\langle \hat M_{2}\rangle_{w}$. The dashed curves represent theoretical
    prediction, while the symbols are the experimental values. ({\bf a}) At  the post-selective state  $|+m_{3}\rangle$. ({\bf b}) At  the   post-selective state  $|-m_{3}\rangle$.   Obviously, the anomalous weak values exist when the post-selection probability is less than half of the maximum value.  Error bars indicate $\pm1$ standard deviation.
    }\label{f2}
    \end{figure}

\begin{figure}[h] \centering
\includegraphics[width=0.5\textwidth]{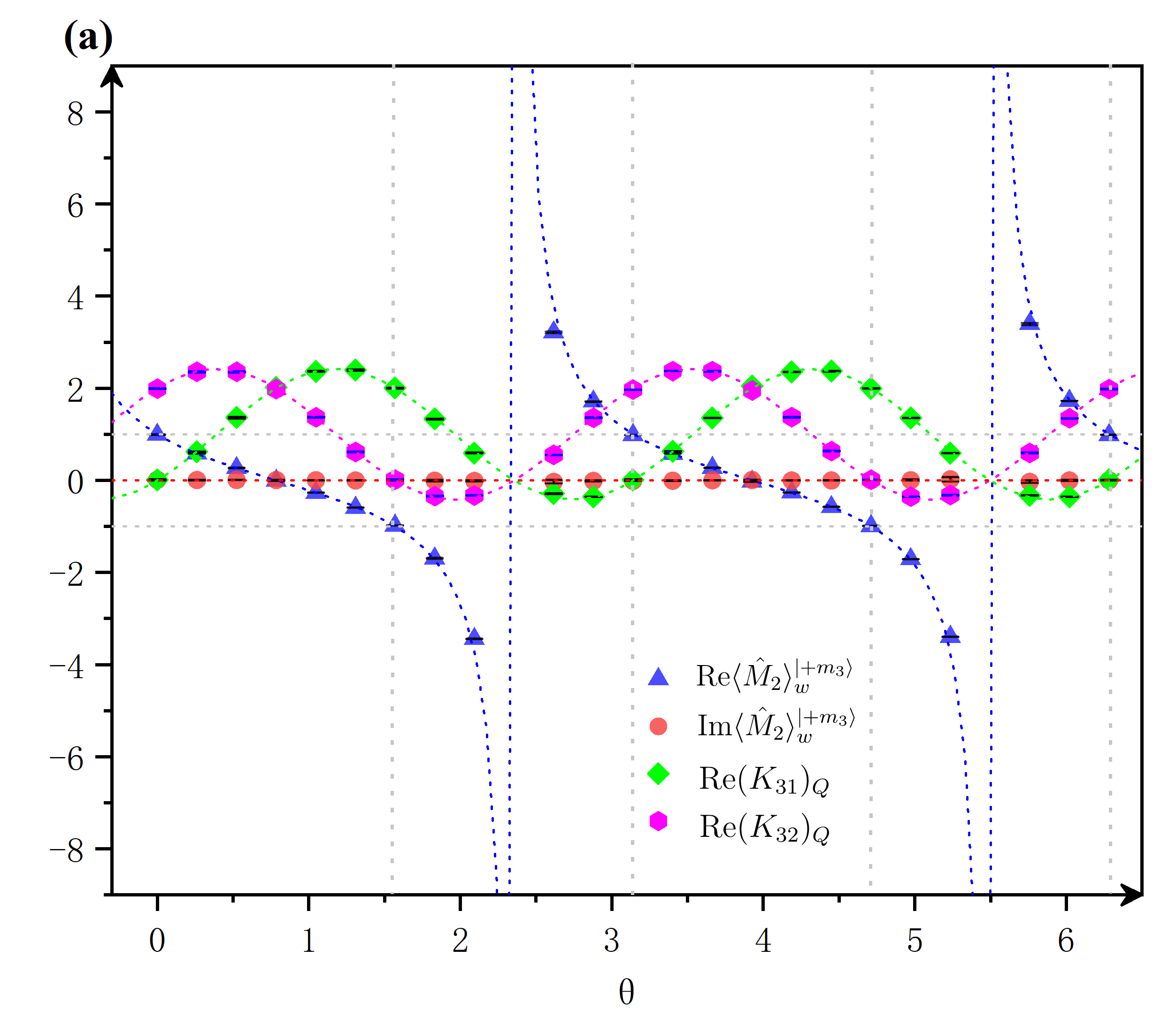}
\includegraphics[width=0.5\textwidth]{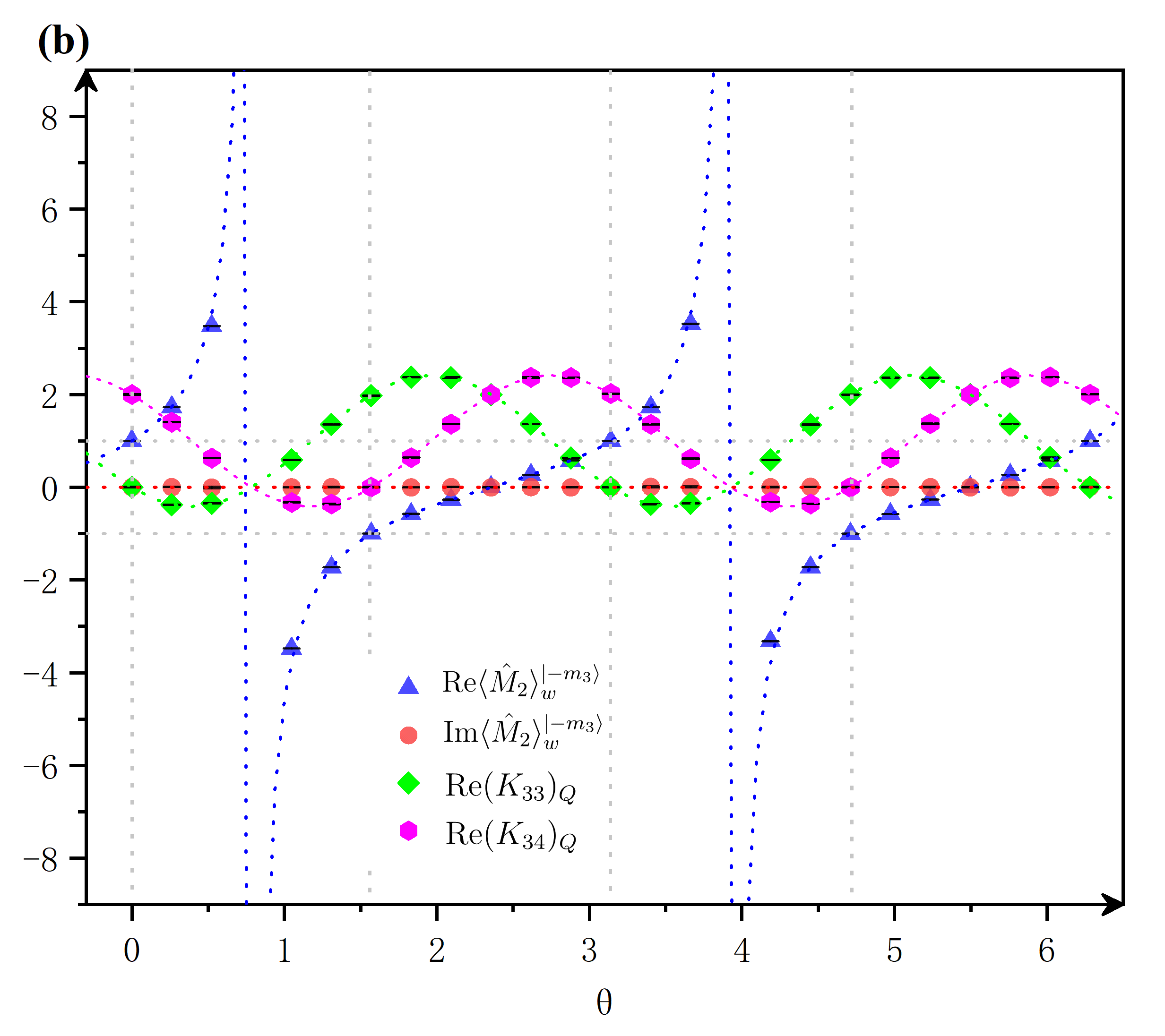}
\caption{The results of the verification of LGIs (\ref{x2}).  ({\bf a}) The experimental values  of $(K_{31})_{Q}$ and $(K_{32})_{Q}$ with post-selective state  $|+m_{3}\rangle$. ({\bf b}) The experimental values  of $(K_{33})_{Q}$ and $(K_{34})_{Q}$ with  post-selective state  $|-m_{3}\rangle$.  The   symbols and dashed curves represent the experimental values and theoretical predictions. There is a violation of LGI in the range of anomalous weak values. Conversely,  violation determine that the weak value is anomalous.  Error bars indicate $\pm1$ standard deviation.
}\label{f3}
\end{figure}

According to  Eq.(\ref{xx2}, \ref{s}),  the weak value $\langle \hat {M}_{2}\rangle_{w}$ and the LGIs (\ref{orgin2})  for quantum systems are calculated based on the light intensity detected by the optical power meter (OPE). And the measurement strength, which is continuously adjustable by parameter $\gamma$, is typically taken at $12^{\circ}$ in Fig. \ref{f2} and Fig. \ref{f3}.

Figure \ref{f2}  show  the experimental results   of the real and imaginary part of  weak value as well as the post-selection probability. The experimental imaginary part $\operatorname{Im}\langle\hat{M}_{2}\rangle_{w}^{\left|\pm m_{3}\right\rangle}$ is nearby-zero, which is consistent with the theoretical prediction that the weak value is real.  When the   post-selection probability satisfy $p_{i}< 0.5$, the real  part is   either $\operatorname{Re}\langle\hat{M}_{2}\rangle_{w}^{\left|\pm m_{3}\right\rangle}>1$ or $\operatorname{Re}\langle\hat{M}_{2}\rangle_{w}^{\left|\pm m_{3}\right\rangle}< -1$.
  That is to say, it will be accompanied by an anomalous weak value if there is destructive   interference at the detection port.
We noticed that the postselection probability can be amplified by $N$ times  while the weak value remains fixed \cite{chenjs1,Kim2022}.
For example, the dashed cyan and magenta  lines in Fig.  \ref{f2}.({\bf a}) correspond to magnification  $N=2$   and $N=3$, denoted by $p_{|+m_{3}\rangle}^{N=2}$ and $p_{|+m_{3}\rangle}^{N=3}$, respectively. It also can be seen that an anomalous weak value exist once  the post-selection  probability is less than half of the maximum value.

The corresponding  results of  LGIs (\ref{x2}) are shown in Fig. \ref{f3}, where the dashed  curves are the theoretical prediction and the  symbols are the experimental values. Fig. \ref{f3}({\bf a}) and \ref{f3}({\bf b}) display the values of $(K_{3i})_{Q}     (i=1,2$ and $3,4)$, $\operatorname{Re}\langle \hat M_{2}\rangle_{w}$ and $\operatorname{Im}\langle \hat M_{2}\rangle_{w}$ for post-selective state $|+m_{3}\rangle$ and $|-m_{3}\rangle$, respectively.  There is definitely an anomalous weak value  when
one of the equations satisfies  $(K_{3i})_{Q} < 0$, since LGIs (\ref{orgin1})
cannot  hold simultaneously. Conversely, the existence of anomalous weak value certainly indicates the violation of a specific LGI. Conjoined with the results of Fig. \ref{f2}, it suggests that  whenever the destructive effect is observed in the interference, anomalous weak values exist and the  violation of  LGI is further obtained. These values thus show clear experimental evidence that
 the macrorealism is verified by using  interference experiment with weak measurement.

\begin{figure} [thb]\centering
    \includegraphics[width=0.45\textwidth]{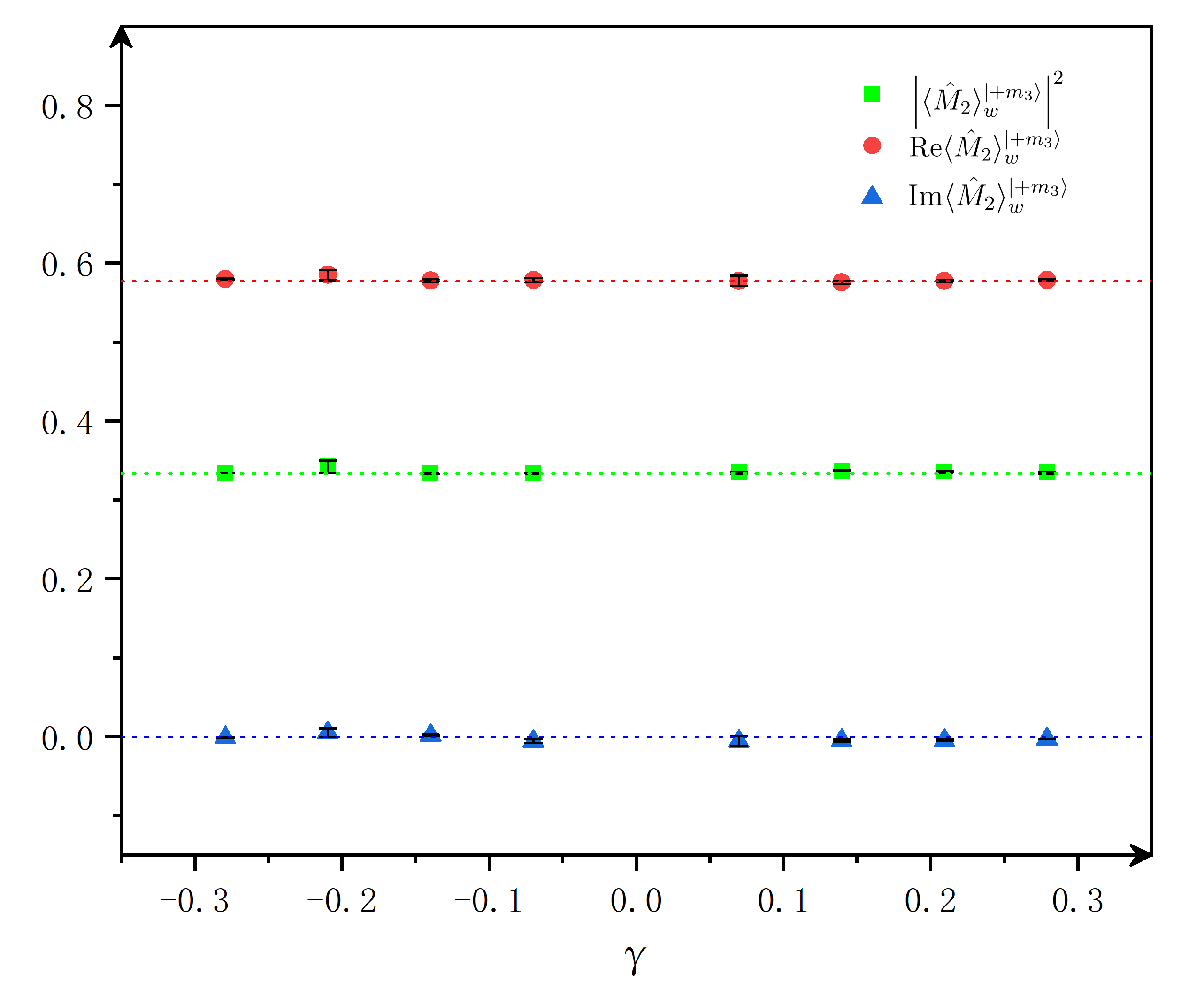}
    \caption{Experimental results of weak value $\langle\hat{M}_{2}\rangle_{w}^{|+ m_{3}\rangle}$ at  varying  coupling strength $\gamma$. The dashed green, blue and red curves represent the norm, imaginary and real
    part of theoretical   values, labeled as  $|\langle\hat M_{2} \rangle_{w}^{|+ m_{3}\rangle}|^{2}$,  $\operatorname{Re}\langle \hat {M}_{2}\rangle_{w}^{|+ m_{3}\rangle}$ and $\operatorname{Im}\langle \hat {M}_{2}\rangle_{w}^{|+ m_{3}\rangle}$. The  symbols represent experimental values. It is evident that the weak value  are independent of coupling
    strength. Error bars indicate $\pm1$ standard deviation.}\label{f7}
    \end{figure}

On the other hand, to verify the dispensability of NIM, different coupling strengths suffered by the measured system  need to be allowed.
Therefore, we set the pre- and post-selection states as $\left|\psi_{i}\right\rangle=(0.966 \left|U\right\rangle + 0.259 \left|D\right\rangle) $, $\left|\psi_{f}\right\rangle=\left|+ m_{3}\right\rangle$.
 Fig. \ref{f7} shows the experimental obtained  values of  $|\langle\hat M_{2} \rangle_{w}^{|+ m_{3}\rangle}|^{2}$,  $\operatorname{Re}\langle \hat {M}_{2}\rangle_{w}^{|+ m_{3}\rangle}$ and $\operatorname{Im}\langle \hat {M}_{2}\rangle_{w}^{|+ m_{3}\rangle}$  for different measurement strength $\gamma$ in the range of $24^{\circ}$ to $-24^{\circ}$ with the step of $4^{\circ}$. The weak value keep constant with the varying  strength, which is in accordance with its  definition.  In fact, the experiment allows setting arbitrary coupling strengths even infinitely close to zero, while do not altering the results. Because the Eq. (\ref{xx2}) for two-dimensional observable $\hat {M}_{2}$ is obtained without approximation. That is a favorable condition, compared with the previous weak measurement method that required the perturbation to be within the linear-response regime \cite{review2012,review2014}.  Such an advantage moves further toward the initial purpose of the NIM assumption, resulting in a direct  test of macroscopic realism.

 The error bars in  figures are calculated by the error propagation formula.
Considering the imperfectness of calibration  or the slowly drift and slight vibrating of the interferometer in the measurement duration,  the corresponding experimental data   agrees with the theoretical prediction.

 \section{ Conclusion}

 In this work, we have  experimentally demonstrated the violation of the Leggett-Garg inequality for path observable   in a composite interference network, which consist of polarization and path degrees of freedom.
Results confirm that the existence of destructive interference inherently implies a violation of macro-realism. Furthermore, the present experiment   complements the loopholes of previous ones \cite{c2021} with precise weak measurement, i.e., the realism  has  convincingly demonstrated in the macroscopic scale using classical coherent light.
 Another salient feature is that results of LGI violation always hold and remain constant even if a perturbation of arbitrary strength is introduced into the system. Our tests therefore achieve a direct measurement whereby the violation of a macrorealist inequality is only caused by the violation of realism per se, releasing the requirement of the NIM  assumption. This may not only provoke deep thought on the relation between macroscopic coherence and quantum fundamental issue, but also enrich the applications in quantum information processing.

\section*{Acknowledgements}
\noindent
This work was supported in part by the National Natural Science Foundation of China(NSFC) under the Grants 11975236, 12235008 and by the University of Chinese Academy of Sciences.

\end{document}